\documentclass[aps,twocolumn,epsfig]{revtex4}
\usepackage{graphicx} 
\usepackage{psfrag} 
\begin{document}  
\title{Geometrical indications of adsorbed hydrogen atoms on 
graphite producing starlike and ellipsoidal features in 
scanning tunneling microscopy images}
	
\author{Mohammad Khazaei,
 Ahmad Ranjbar, Mohammad Saeed Bahramy, 
Hiroshi Mizuseki, and Yoshiyuki Kawazoe}

\affiliation{Institute for Materials Research, 
Tohoku University, Sendai 980-8577, Japan}

\date{\today}

\begin{abstract}

Recent scanning tunneling spectroscopy (STM) experiments 
display images with star and ellipsoidal like features 
resulting from unique geometrical arrangements of 
a few adsorbed hydrogen atoms on graphite. Based on first-principles STM simulations, 
we propose a new model 
with three hydrogen atoms adsorbed 
on the graphene sheet in the shape of 
an equilateral triangle with a hexagon ring surrounded inside. 
The model reproduces the 
experimentally observed starlike STM patterns. 
Additionally, we confirm that an ortho-hydrogen pair is the configuration
corresponding to the ellipsoidal images. These calculations reveal that when 
the hydrogen pairs are in the same orientation, they are energetically more stable.


\end{abstract}

\maketitle
\begin{figure}[t]
\begin{center}
\includegraphics[scale=0.70]{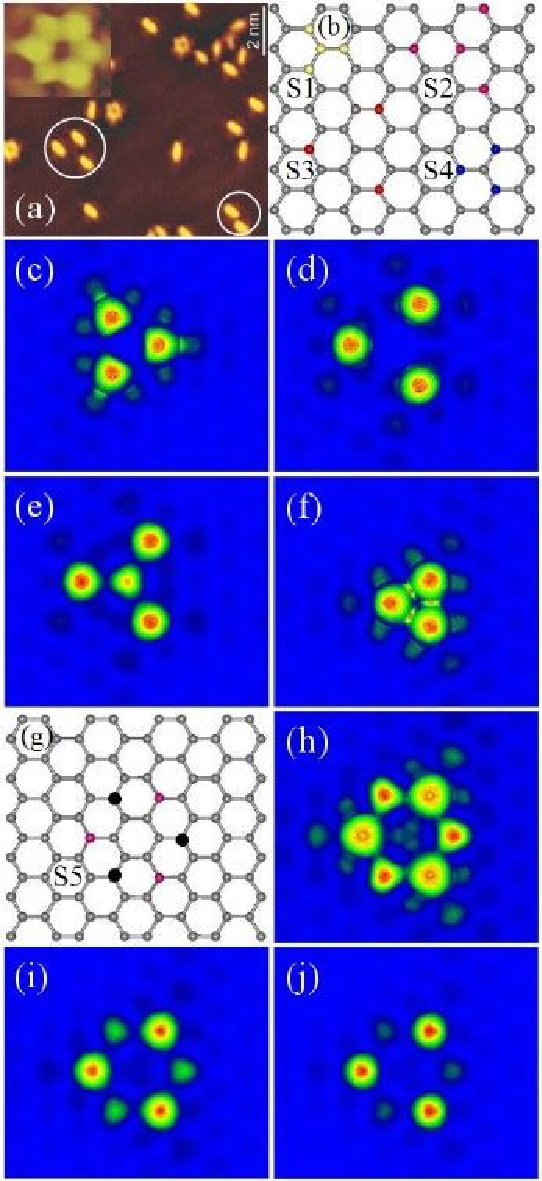}
\end{center}
\caption{(Color online) (a) Experimental STM image of adsorbed 
hydrogen atoms on graphite with
starlike and ellipsoidal features~\cite{hornekar3}. The magnified starlike feature is shown in the inset, (b)
the experimentally proposed configurations for the 
adsorbed hydrogen atoms producing the starlike STM 
images~\cite{hornekar3}, and (c-f) the computed STM 
images for S1-S4 at 0.5 V/\AA, respectively. 
Red (blue) color denotes the highest (lowest) intensity 
of the emitted currents.
(g) Geometrical structure of S5, and (h-j) 
its corresponding STM images computed at 0.7, 0.5, and 0.3 V/\AA, 
respectively.} 
\label{fig:experiment}
\end{figure}

Hydrogen adsorption on graphite surface has been the subject of many theoretical 
and experimental studies,  due to its fundamental importance in both science and technology
~\cite{roman,boukhvalov,ferro,yazyev,lei,hornekar1,hornekar2,hornekar3}. 
The STM technique is known as one of the most powerful 
experimental tools for analyzing surfaces. Through this technique, 
for example, Hornek{\ae}r \textit {et al.} considered the 
atomic structures of various types of H(D) clusters on 
graphite~\cite{hornekar1,hornekar2}. In their latest 
experiments, they identified two types of
 clusters as the most abundant hydrogen species on graphite 
surface. The STM images of these clusters exhibit 
star and ellipsoidal like features, see FIG.~\ref{fig:experiment}(a)~\cite{hornekar3}.   
The former includes six bright spots, three of which  
are relatively larger than the others~\cite{hornekar3}. 
Considering the STM images observed 
at various bias voltages, Hornek{\ae}r \textit {et al.} 
concluded that the starlike feature originates from a particular 
arrangement of three or four hydrogen atoms 
adsorbed on graphite surface~\cite{hornekar3}. 
For the ellipsoidal STM
feature, using a set of STM simulations, Hornek{\ae}r 
\textit {et al.} proposed that 
it should represent a pair of hydrogen atoms adsorbed on 
two adjacent  carbon atoms, so-called ortho-hydrogen pair. 
Although, the computed STM 
image by Hornek{\ae}r \textit {et al.} appears to be
in agreement with experiment, the authors have not clarified 
how a pair of hydrogen atoms with H-H distance $\sim$2.1 \AA~can 
produce an elongated-ellipsoidal feature with a length of
$\sim$7.0 \AA ~\cite{hornekar1,hornekar3}.
In the present work we propose a model whose computed STM image perfectly 
matches  with experimentally observed starlike STM images. Additionally, 
we confirm that the ortho-hydrogen pair is the configuration for the
 ellipsoidal images and explain how these pairs produced the elongated features in STM images. 
We further investigate the interaction of hydrogen pairs together. 
Our calculations show that when the pairs are oriented in the same direction, 
they are energetically more stable.

The electronic structure calculations 
are carried out within the context of density functional 
theory (DFT) using the spin-polarized 
Perdew$-$Burke$-$Ernzerhof (PBE) exchange-correlation 
functionals and the projected augmented wave (PAW) method, 
as implemented in the VASP code~\cite{vasp}.  
The adsorbed hydrogen atoms on graphite surfaces are 
simulated using a graphene sheet  
containing 96 carbon and appropriate number of hydrogen atoms (two, three, or four). 
The corresponding Brillouin 
Zone is sampled by a 5$\times$5$\times$1 Monkhorst-Pack mesh. 
The structures are fully optimized until the magnitude of 
force on each ion becomes less than 0.04 eV/\AA. There 
is currently no simple way to apply the electric field in the VASP 
calculations. Accordingly, we use the SIESTA, which is a 
DFT code with localized basis sets~\cite{soler}, for 
the simulation of STM images. 
We performed a new set of spin-polarized single-point energy 
calculations for the structures, previously optimized by VASP, 
in the same level of theory  (DFT/PBE) 
but at definite electric fields. 
A full description of our methodology for STM 
calculations can be found in Ref.~\cite{khazaei1,khazaei2,penn}.

To identify the geometrical positions of hydrogen atoms on 
graphite producing the starlike STM image, we first examine 
models with four H atoms (labeled as S1 and S2 ) and three H 
atoms (labeled as S3 and S4), proposed by Hornek{\ae}r 
\textit{et al.}~\cite{hornekar3}, see 
FIG.~\ref{fig:experiment}(b). Figures~\ref{fig:experiment}(c-f) 
show the STM images computed for S1, S2, S3, and S4 
under an electric field of 0.5 V/\AA. 
From FIGs.~\ref{fig:experiment}(c-f), it is evident that none of the above 
models can reproduce the experimentally observed STM pattern. 
Interestingly, the computed STM images of S1 and S2 with four 
hydrogen atoms have only three bright spots while the S3 
structure with three hydrogen atoms shows 
four bright spots in its STM pattern. 
This can be attributed to the fact that the computed current is 
substantially proportional to the electron tunneling and LDOS. 
For S1 and S2, our calculations 
show almost no LDOS at left turning points of the central 
hydrogen atom. As a result, 
the emitted current from the central 
hydrogen atom becomes negligibly small. On the other hand, 
for S3, both the electron tunneling and LDOS turn out to be 
significantly higher for H atoms and their central carbon atom. 
Accordingly, these four atoms contribute substantially to the total emitted 
current, as can be seen in FIG.~\ref{fig:experiment}(e).

We propose a new model S5, in which three hydrogen atoms 
are symmetrically placed on the graphene sheet in an
 equilateral triangle encompassing a complete hexagon ring of carbon atoms, 
 see FIG.~\ref{fig:experiment}(g).  The configuration S5 has a large magnetic moment of 3$\mu_B$.

The spin-polarized calculation shows that total energy is approximately 
0.1 eV less than that estimated by non spin-polarized calculation. Therefore the S5 
structure is expected to be magnetic even at room 
temperature. 
Figures~\ref{fig:experiment}(h-j) 
show the STM images of S5 computed at 0.7, 0.5, and 0.3 V/\AA, 
respectively. The computed STM images all show a starlike 
feature similar to that observed in experiment. There are 
six bright spots, three of which look relatively larger than the 
other three. Comparing the position of the spots 
and the geometrical configuration of S5, the centers of the 
larger (smaller) spots turn out to be on the hydrogen atoms 
(the carbon sites indicated by black spots in FIG.~\ref{fig:experiment}(g)). 
In agreement with experimental results reported by Hornek{\ae}r 
\textit{et al.}, our simulations demonstrate that when the strength 
of applied voltages decreases (here from 0.7 to 0.3 V/\AA), 
the intensity of the small spots reduces. To explain the low intensity of 
small spots at low voltages, we have considered the patterns of total LDOS 
and total tunneling probability of S5 (not shown here). These results  
show that the LDOS patterns are almost the same under various applied voltages. 
At high-applied voltage 0.7 V/\AA, the tunneling probabilities of 
electrons from both hydrogen and carbon atoms are large. At lower 
voltages, 0.3 and 0.5 V/\AA, the tunneling probability of electrons 
from hydrogen atoms is relatively larger than that from carbon atoms.  
Such a difference  can be attributed to the 
fact that the adsorbed hydrogen atoms create dipole moments on the 
graphene surface whereby the effective potential around the hydrogen 
atoms decreases and consequently, the tunneling probability of electrons 
from them increases. This explains the reason for high electron emission from 
hydrogen atoms in all ranges of applied voltages. 
It is worth mentioning that further calculations using a larger unit cell with a side length of $\sim$35\AA~ reveals no change in the computed STM images and magnetic properties of S5.

\begin{figure}[t]
\begin{center}
\includegraphics[scale=0.37]{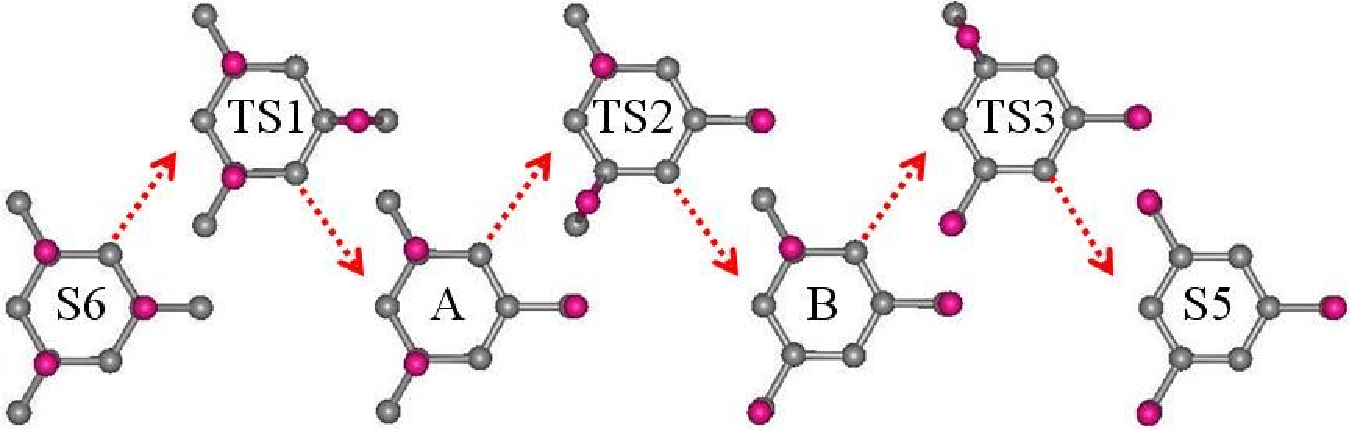}
\end{center}
\caption{(Color online) The geometrical configuration of S6 and its 
possible transformation pathway to S5.} 
\label{fig:jump}
\end{figure}

To compare the stability of S5 with other structures, we summarize in 
Table~\ref{tab:energy}, the respective adsorption energy values of hydrogen atoms 
for optimized S3, S4, S5 and S6 configurations (the latter is illustrated in FIG.~\ref{fig:jump}). 
The adsorption energy is defined as $E_a$=$E_{graphene+nH}-E_{graphene}-nE_H$ where $E_{graphene+nH}$, $E_{graphene}$ and $E_H$ 
are the total energies obtained for a graphene sheet with n adsorbed hydrogen atoms, a perfect graphene sheet and an isolated H atom, respectively. 
Overall comparison shows that the most stable structures, S3 and S5, 
are energetically isomeric. However, since the STM image of S3 is totally different 
from that observed in experiment,  S5 appears to be  
practically preferable to S3. This implies that there should be a post-adsorption mechanism affecting the structure of initially adsorbed hydrogen clusters on graphite, 
otherwise, S3 and S5  should statistically have had 
the same population, in practice. 
Thus, the experimental conditions, within which the above STM images have been observed, 
need to be carefully taken into account. In the experiment, S5 is observed in 
samples which are annealed to 525K 
after a heavy  deposition of H(D) atoms on the surface. Since
the utilized H(D) beam is extremely hot (1600-2200 K),  hydrogen clusters, in various configurations 
including S1-S6, are created~\cite{ hornekar2,hornekar3}. 
However, by annealing the sample, many of these clusters are evaporated 
from the surface or change to other configurations. A statistical study by 
Hornek{\ae}r \textit{et al.} reveals that, the relative abundance  of 
two specific H(D)  species  with starlike and ellipsoidal STM 
features  dominantly increases, as the sample is further annealed to 570 K~\cite{hornekar3}. 
The process, described above, suggests that after annealing the sample, many H(D) species initially deposited on the surface  transform into S5 configuration.  
For such  a transition, basically, species with similar symmetry but higher energy such as S6  are preferred. 
\begin{table}
\caption{Adsorption energy (E$_{a}$) values obtained for the  triatomic
configurations, S3-S6, and the  dimer configurations, D1-D7.}
\begin{tabular}{cccc}
\hline
\hline
structures &  E$_{a} (eV)$  & structures  &  E$_{a}$ (eV) \\               
 \hline
   S3   &    -2.63    & D1    & -2.8    \\
  S4    &    -2.61       &D2   &    -1.65   \\
  S5    &   -2.62        & D3  &    -2.75     \\
  S6    &   -2.43        &D4    &  -1.59      \\
        &                  &D5      & -2.18      \\
        &                  & D6      &   -1.67      \\
        &                  & D7      & -1.83    \\
\hline
\hline
\end{tabular}
\label{tab:energy}
\end{table}

A possible transition pathway from S6 to S5  is  depicted in FIG.~\ref{fig:jump}.  In this picture, The diffusion reactions S6$\rightarrow$A, A$\rightarrow$B, and B$\rightarrow$S5 are assumed to pass through the barriers $\Delta E^\star_1=E^\star_{TS1}-E^\star_{S6}$,  $\Delta E^\star_2=E^\star_{TS2}-E^\star_{A}$ and $\Delta E^\star_3=E^\star_{TS2}-E^\star_{B}$ whose saddle points are at the transition states, TS1, TS2, and TS3, respectively. Note that, above all $E^\star$ terms include the zero-point energy correction. To find the correct transition paths and the corresponding energy barriers, we use  the nudged elastic band method  ~\cite{henkelman}. As a criterion for determining the saddle points, the  phonon eigenmodes of each TS are examined so that they have one and just one imaginary frequency. Accordingly, $\Delta E^\star_1$, $\Delta E^\star_2$ and $\Delta E^\star_3$, are obtained to be 0.48 eV, 0.84 eV and 0.86 eV, respectively. The maximum influence of vibrational  zero-point energy correction on barriers is less than 0.12 eV.  For the sake of comparison, we have also calculated the energy barrier for desorbing each of hydrogen atoms, supposed to be displaced through S6$\rightarrow$A, A$\rightarrow$B, and B$\rightarrow$S5 reactions. The respective desorption energies are  0.67, 1.0, and 1.2 eV. Thus by annealing, the hydrogen atoms in S6 prefer to diffuse to S5 through the  process described above rather than to be desorbed from the graphene surface. 

 \begin{figure}[t]
\begin{center}
\includegraphics[scale=0.3]{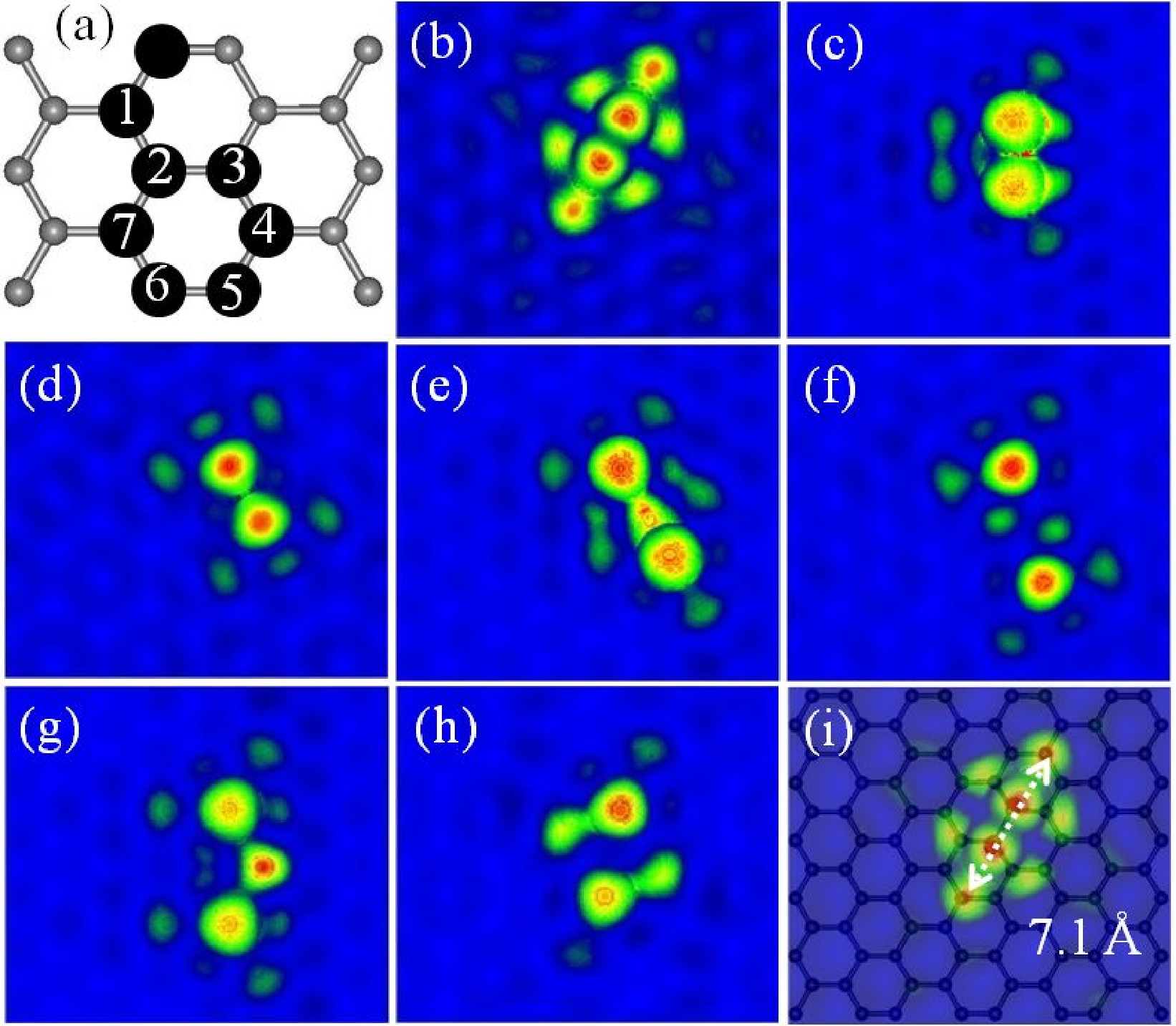}
\end{center}
\caption{(Color online) (a) the geometrical structure of D1-D7, (b-h) 
the respective STM images computed for D1-D7. (i) The projection of 
STM image of D1 onto its geometrical configuration.} 
\label{fig:dimers}
\end{figure}

To further asses the possibility of above reactions at different temperatures, we have estimated the rate of reactions at 300K and 500K within the transition-state theory~\cite{vegge}. In this approach, the rate of reaction is defined as $r$=$\frac{\Pi_i 1-\exp(-\hbar\omega_i^{IS}/k_BT)}{\Pi_i' 1-\exp(-\hbar\omega_i^{TS}/k_BT)}\exp\left(\frac{\Delta E^\star }{k_BT}\right)$ where $\omega_i^{IS}$ and  $\omega_i^{TS}$  are the eigenmodes of initial state (IS) and transition state (TS), and as pointed out earlier $\Delta E^\star$ is the energy difference between TS and IS after including zero point energy correction. On this basis, the respective values of $r$ at 300K and 500 K are expected to be
S6$\rightarrow$A: $11.2\times10^4s^{-1}$ 
and $3.1\times10^8s^{-1}$, A$\rightarrow$B: $6.0\times10^{-2}s^{-1}$ 
and $4.3\times10^4s^{-1}$, 
and B$\rightarrow$S5: $2.7\times10^{-2}s^{-1}$, 
and $2.8\times10^4s^{-1}$.  
These results clearly verify that the proposed transition process is practically very feasible. It is necessary to note  that  S5 may not be the most stable configuration for adsorption of three hydrogen atoms on graphene. Rather, it is a metastable structure made in a particular experimental conditions.

As mentioned earlier, the experimentally observed STM images include 
patterns with elongated-ellipsoidal features ~\cite{hornekar3}. 
Hornek{\ae}r \textit {et al.} proposed that such features originate 
from a pair of H atoms, adsorbed on two adjacent carbon atoms of graphite~\cite{hornekar1}. 
Although their model, here labeled as D1, 
appears to reproduce the same ellipsoidal STM features, their calculations fail to explain, 
how a pair of hydrogen atoms with H-H 
distance of $\sim$2.1\AA~can produce a bright ellipsoidal feature 
as long as $\sim$7.0\AA~\cite{hornekar3}. To answer this question, 
we have computed the STM image for D1 and other possible H dimers. 
The structure of dimers and their corresponding STM images are 
illustrated in FIG.~\ref{fig:dimers}. Additionally, in 
Table~\ref{tab:energy}, we have summarized the values of 
adsorption energies. According 
to FIG.~\ref{fig:dimers}, each dimer has a unique STM pattern. Reassuringly, 
 D1, is the only dimer structure whose STM 
image represents ellipsoidal feature, similar to that observed in 
experiment. Interestingly, the bright ellipsoidal pattern obtained in our STM 
calculation for D1 has almost the same experimental length $\sim$7.0\AA. 
For the sake of clarity, in FIG.~\ref{fig:dimers}(i) we have projected the D1's STM 
image onto its geometrical structure. Evidently, not only 
the hydrogen atoms but also their neighboring carbon atoms contribute 
significantly to the total STM current. In other words, the creation of D1 on 
graphene changes the electronic structure of its surrounding carbon atoms  
such that their corresponding LDOS values  become significantly large. Consequently, 
they contribute substantially  to the STM current. D1 is nonmagnetic system and according to Table~\ref{tab:energy}, 
it is energetically the most stable configuration in comparison to other dimer models.  Experimental 
results also show that D1 is the most abundantly formed configuration. 

\begin{figure}[t]
\begin{center}
\includegraphics[scale=0.57]{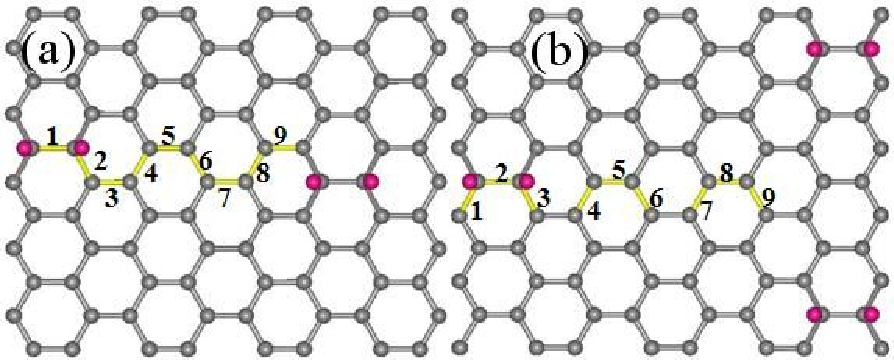}
\end{center}
\caption{(Color online) The relative positions of (a) two hydrogen dimers and (b) three hydrogen 
dimers on a graphene sheet. The highlighted C-C bond denoted by numbers 1-9 indicate the position of displaced dimer.} 
\label{fig:ttd}
\end{figure}

Interestingly, it appears that many of the H dimers absorbed on graphite, seems to be oriented in the same direction. This can be clearly seen in FIG.~\ref{fig:experiment}(a), see the bright spots discriminated by the white circles. 
To elaborate on this observation, we have carried out a set of calculations in which the adsorption energy is calculated for two and  three H dimers when they have different orientations and distances in respect to each other. In our models as shown in FIGs.~\ref{fig:ttd} (a) and (b), we keep the position of one or two of dimers fixed on the surface, while the other dimer is displaced so that for each calculation it is on one of C-C bonds, indicated by numbers 1-9. To minimize the interaction of dimers with their periodic images in the neighboring unit cells, for calculations with two (three) H dimers we have considered a very large unit cell 34.08\AA$\times$14.76 \AA~ (34.08\AA$\times$29.52 \AA), containing 192 (384) C atoms. The corresponding values of adsorption energies are summarized in Table~\ref{tab:adsorption}. The results clearly indicate that dimers with same direction become energetically more stable, as they get closer to each other. On the other hand, the disorientation of H dimers results in an increase in the surface energy and, hence, in instability of whole structure. Consequently, the adsorbed H dimers prefer to diffuse on graphene surface so that they can stay in the same directions.  

\begin{table}[t]
\caption{Adsorption energies E$_{a}$ of hydrogen dimers on graphene sheets shown in 
FIGs.~\ref{fig:ttd}(a) and (b)}
\begin{tabular}{ccc}
\hline
\hline
Position of &  E$_{a}$ of two dimers &  E$_{a}$ of three dimers\\        
 displaced dimer &   (eV) &   (eV) \\         
 \hline
   1   &  -5.58      &  -8.18     \\
  2    &  -5.49      &  -8.21     \\
  3    &  -5.69      & -8.19      \\
  4    &   -5.54     &   -8.18    \\
   5     & -5.66      &  -8.22    \\
   6     & -5.39      &  -8.16    \\
   7     & -6.17      &  -8.12    \\
   8     & -5.64      &  -8.27    \\
   9     & -5.71      & -8.18     \\
\hline
\hline
\end{tabular}
\label{tab:adsorption}
\end{table}

In conclusion based on STM image calculations, we have identified the 
geometrical configurations of the most abundant species of adsorbed hydrogen 
atoms on graphite after a heavy dosing of H(D) atoms.  
The structures were shown to have two and three 
hydrogen atoms with STM images having elongated-ellipsoidal 
and starlike features, respectively. 
The former (latter) turned 
out to be nonmagnetic (strongly magnetic). In the case of hydrogen pairs,
they are 
energetically more stable when oriented in the same direction.  

M.K. and H.M. acknowledge their funding from the New Energy and Industrial Technology 
Development Organization (NEDO). The authors are grateful to Prof. M. Philpott 
and Prof. R. V. Belosludov for their helpful comments.


\begin{thebibliography}{99}
  \bibitem[1]{roman}
  T. Roman \textit {et al.}, Carbon {\bf 45}, 203 (2007).
  \bibitem[2]{boukhvalov}
D. W. Boukhvalov \textit {et al.}, Phys. Rev. B {\bf 77}, 035427 (2008).
 \bibitem[3]{ferro}
  Y. Ferro \textit {et al.}, Phys. Rev. B {\bf 78}, 085417 (2008).
 \bibitem[4]{yazyev}
 O. V. Yazyev and L. Helm, Phys. Rev. B {\bf 75}, 125408 (2007).
 \bibitem[5]{lei}
Y. Lei \textit {et al.},
Phys. Rev. B {\bf 77}, 134114 (2008).
\bibitem[6]{hornekar1} 
L. Hornek{\ae}r \textit {et al.}, Phys. Rev. Lett. {\bf 96}, 156104 (2006).
\bibitem[7]{hornekar2} 
L. Hornek{\ae}r \textit {et al.}, Phys. Rev. Lett. {\bf 97}, 186102 (2006).
\bibitem[8]{hornekar3} 
L. Hornek{\ae}r \textit {et al.}, Chem. Phys. Lett. {\bf 446}, 237 (2007).
\bibitem[9]{vasp} 
G. Kresse and J. Furthm\"{u}ller, Comput. Mater. Sci. {\bf 6}, 15 (1996). 
\bibitem[10] {soler}
J. M. Soler \textit{et al.},  
J. Phys.: Condens. Matter {\bf 14}, 2745 (2002).
\bibitem[11]{khazaei1} 
M. Khazaei \textit{et al.}, Phys. Rev. Lett. {\bf 95}, 177602 (2005).
\bibitem[12]{khazaei2} 
M. Khazaei \textit {et al.}, J. Phys. Chem. C {\bf 111}, 6690 (2007).
\bibitem[13]{penn}
D. R. Penn and E. W. Plummer, Phys. Rev. B {\bf 9}, 1216 (1974). 
 \bibitem[14]{andree} 
A. Andree \textit {et al.}, Chem. Phys. Lett. {\bf 425}, 99 (2006).
 \bibitem[15]{henkelman}
 G. Henkelman  \textit {et al.}, J. Chem. Phys. {\bf 113}, 9901 (2000). 
 \bibitem[16]{vegge}
T. Vegge, Phys. Rev. B {\bf70}, 035412 (2004). 
\end{thebibliography}
\end{document}